\documentclass{elsart}
\usepackage{epsf}

\begin{document}
\begin{frontmatter}
  \title{Coherent and Incoherent structures in systems described by
  the 1D CGLE:\\ Experiments and Identification} \author[rul]{Martin
  van Hecke} \address[rul]{Kamerlingh Onnes Laboratorium, Leiden
  University,\\ Niels Bohrweg 2, 2333CA, Leiden, The Netherlands}

\begin{abstract}
Much of the nontrivial dynamics of the one dimensional Complex
Ginzburg-Landau Equation (CGLE) is dominated by propagating structures
that are characterized by local ``twists'' of the phase-field. I give a
brief overview of the most important properties of these various
structures, formulate a number of experimental challenges and address
the question how such structures may be identified in experimental
space-time data sets.


\end{abstract}
\end{frontmatter}

\section{Introduction}

A large body of theoretical work has been devoted to unravel the
intricate space-time dynamics of the 1D CGLE. One of the most exciting
developments over the past few years has been the understanding of
various aspects of the non-trivial, fully nonlinear behavior of this
model in terms of the properties of a number of ``coherent
structures''.  While much earlier work has focused on so-called
Nozaki-Bekki (NB) holes \cite{nb}, two other, recently revealed
families of coherent structures appear to play the dominant role in
large parts of parameter space. These structures are the weakly
nonlinear {\em Modulated Amplitude Waves} (MAWs) that occur when plane
waves become linearly instable \cite{maw,lutznew} and the related, but
more nonlinear, {\em Homoclons} which are connected to
defect\footnote{The defects that occur in the 1D CGLE are sometimes
referred to as phase-slips; these defects do not persist, but rather
occur at isolated points in space-time.}  formation
\cite{clon,mm,zz}. The work on these two structures has been published
rather recently and it is therefore no surprise that, at present,
their experimental relevance is unclear.

Coherent structures have a fixed spatial profile, and their dynamics
is a combination of propagation and oscillation; in CGLE-language,
$A(x,t)=\exp^{i -\omega_{coh} t} \tilde{A}( x -v_{coh}t)$.  One would
expect that coherent structures observable in experiments are
{\em{(i)}} linearly stable (such that they are attracting) and
{\em{(ii)}} structurally stable (such that small perturbations of the
equations of motion do not destroy them). If a certain coherent
structure is (structurally) unstable, one expects to see, in
experiments or simulations, that the ``shape'' of the structure
changes over time: I will refer to such structures as {\em incoherent}
structures. One may require that by a smooth change of parameters or
initial conditions, a certain incoherent structure can be brought
arbitrarily close to its coherent counterpart\footnote{this is more
difficult for structural instabilities, which are due to perturbations
of the underlying dynamical system that may not so easily be
reversed}; the slower the nontrivial dynamics is, the ``more
coherent'' a certain structure is.

The crucial complication one encounters when confronting theoretical
predications for the behavior of local structures to real experimental
data is that Homoclons and MAWs are (almost) always linearly unstable
\cite{maw,lutznew} (see also section \ref{limmaw}), while almost all
NB holes are structurally unstable \cite{nbunst}.  Nevertheless, the
unstable structures are important building blocks for the dynamics of
the CGLE \cite{maw,lutznew,clon,mm,zz}; in many states {\em
incoherent} structures occur, and these are often related to the
unstable {\em coherent} Homoclons and MAWs. Therefore the study of
experimental {\em space-time} data sets is, I believe, essential for
the understanding of 1D wave systems: snapshots of the field simply do
not contain enough information \cite{chate}.  In that sense, the
situation in one dimension is more difficult than in two dimensions,
where a central role is played by spirals that can easily identified
in snapshots of the field \cite{spiral}.

Some additional problems one encounters when comparing experimental
data to theory are: {\em{(i)}} Many of the theoretical studies have
focussed on the chaotic regimes, giving the false impression that MAWs
and Homoclons only play a role in chaotic states.  {\em{(ii)}} In most
experiments the values of the linear and nonlinear dispersion
coefficients, which are essential in the theoretical description, are
not known. {\em{(iii)}} There is some confusion, I believe, concerning
the relevance of so-called Nozaki-Bekki holes \cite{nb}. Since their
analytical form has been known for more than 15 years, these holes
have been studied extensively \cite{nbunst,nbint,review} and there are
some claims in the literature that these are observed in experimental
systems \cite{lega,fleselles,burg}; as I will argue below, it may be
beneficial to have a second look at some of the data.  To clarify this
situation, I will suggest a number of experiments designed to probe
the relevance of MAWs and Homoclons in traveling wave systems. In
addition I will discuss how to distinguish incoherent Homoclons and NB
holes.

The outline of the paper is as follows. In section \ref{seccgle} a
brief theoretical introduction to the CGLE and the coherent structures
framework is given, and the properties of the following coherent
structures and ODE orbits are discussed: (i) Plane waves
(corresponding to fixed points) (ii) Homoclons (corresponding to
homoclinic orbits) (iii) Modulated Amplitude Waves (MAWs) (limit
cycles) (iv) Nozaki-Bekki (NB) Holes (heteroclinic orbits). In Section
\ref{secexp} I give an overview of the dynamical behavior of the
incoherent MAWs and Homoclons and suggest a number of experiments to
probe their properties.  Section \ref{secout} contains conclusions and
a short outlook.

\section{The 1D CGLE}\label{seccgle}

\subsection{Basic properties}

The one-dimensional complex Ginzburg Landau equation describes pattern
formation near a supercritical Hopf bifurcation \cite{review,ch}.  The
focus of this paper is on experiments that produce a single 1D
traveling wave (or a uniform oscillation) via a forward Hopf
bifurcation. In many of such wave systems, states with both left and
right traveling waves occur, and an overview of the behavior of the
coupled Ginzburg-Landau equations that describe this system can be
found in \cite{mysource}. Here it is supposed that the system can be
manipulated so as to contain a single traveling wave state, such that
the 1D CGLE is the appropriate amplitude equation. In its full
dimensional form this equation contains a large number of
coefficients, most of which can be scaled out for a theoretical
analysis (see appendix \ref{app}). After such a rescaling, the CGLE
reads:
\begin{equation}
\partial_t A = A + (1+ i c_1) \partial_{xx} A -(1- i c_3) |A|^2 A~.
\label{cgle}
\end{equation}

The CGLE displays a wide range of behavior as function of the
coefficients $c_1$ and $c_3$. For example, when $c_1$ and $c_3$ are of
opposite sign, the dynamics of the CGLE is essentially relaxational,
while when $c_1$ and $c_3$ go to $\infty$, it is integrable
\cite{review,ch}. Away from these limits, the dynamics interpolates
between ordered and chaotic \cite{shrai,chat2}.

An important symmetry of the CGLE is its phase invariance; if $A$ is a
solution to the CGLE, so is $A e^{i \phi}$ (for constant $\phi$). This
symmetry is related to invariance of the underlying system with
respect to shifts in (space-)time. Writing $A$ in its polar
representation as $A(x,t)= a(x,t) e^ {i\phi(x,t)}$, only the
derivatives of the complex phase are relevant, and it is helpful to
think in terms of the modulus $a$ and the phase-gradient $\partial_x
\phi$. This latter phase-gradient is also referred to as a ``local
wavenumber'', denoted by $q$.

The simplest nontrivial solutions to the CGLE are plane waves of the
form
\begin{equation}\label{qwave}
A= \sqrt{1-q_{pw}^2} \exp (i ( q_{pw} x - \omega_{pw} t)),
\hspace{5mm} \omega_{pw} = - c_3 + q_{pw}^2 (c_1 + c_3)
\end{equation}
Note that the local wavenumber of plane waves (\ref{qwave}) is
constant: $q(x,t)=q_{pw}$.  These states are the background for the
dynamics that will described here. A linear stability analysis reveals
that these waves are prone to the Eckhaus instability when
\cite{bf,janiaud,kramer85}
\begin{equation}\label{nonlq}
  q_{pw}^2 > \frac{(1-c_1c_3)}{3-c_1c_3+2c_3^2}~.
\end{equation}
The band of wavenumbers for which plane waves are stable is widest for
$c_1\!=\!c_3\!=\!0$ and shrinks when these coefficients are increased;
when $c_1 c_3 \!>\!1$ there are no linearly stable waves left and
chaos occurs.  

\begin{figure}
  \vspace{-0cm} \epsfxsize=1.\hsize \mbox{\hspace*{-.01 \hsize}
\epsffile{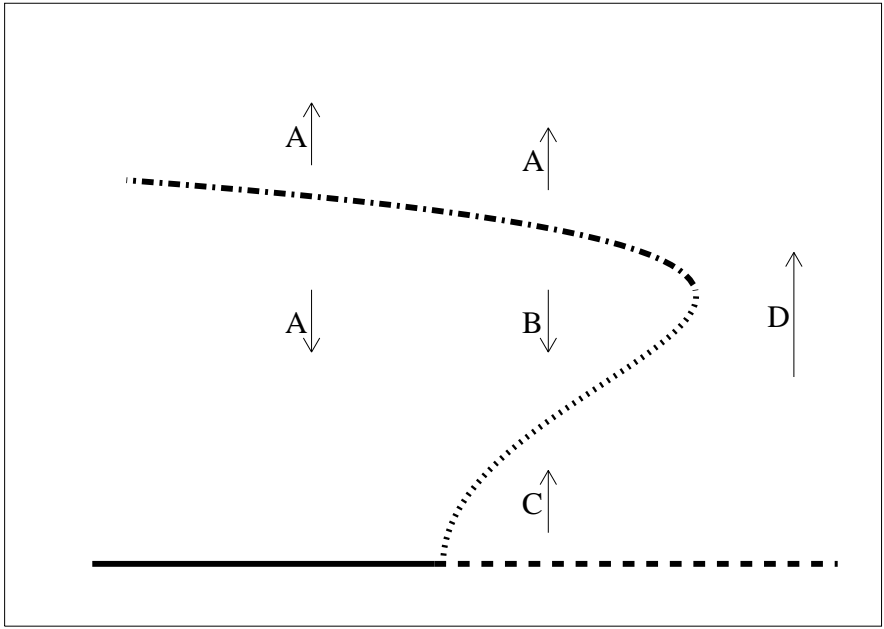} }
\vspace{0.0cm} \caption{Schematic diagram illustrating the relation
between MAWs, Homoclons and CGLE dynamics for a background wave of
small wavenumber. The vertical axis can be thought of as the maximum
phase gradient of a local structure, while the horizontal axis
represents the coefficients $c_1$ and $c_3$.  The curves represent
stable plane waves (continuous line), unstable plane waves (dashed
line), ``Upper branch'' structures (dot-dashed curve) and ``lower
branch'' structures (dotted curve), while the arrows represent typical
incoherent dynamics (see text). Homoclinic and limit-cycle structures
(i.e., isolated holes and periodic modulated structures) occur on both
branches, but the most relevant of these are MAWs, periodic structures
on the lower branch, and Homoclons, isolated structures on the upper
branch. Isolated structures on the lower branch can be called
$P\rightarrow \infty$ MAWs, while periodic structures on the upper
branch can be called ``periodic Homoclons''; neither play an important
role in dynamical states.  For more details see
\cite{maw,lutznew,clon,mm,zz} }
\label{sketch}
\end{figure}

The importance of MAWs and Homoclons can be understood by considering
the dynamical fate of local perturbations of the background wavenumber
of a plain wave consisting of a ``twist'' of the $A$ field of the
CGLE, or, equivalently, a local concentration of $q$. Inside the
stable band, the {\em linear} evolution of the local wavenumber is
given by a combination of diffusion and advection \cite{review}: $q_t
= D q_{xx} + v_{gr} q_x$, where $v_{gr}$ is the {\em nonlinear}
group-velocity: $v_{gr}= \partial \omega_{pw} / \partial q_{pw} =
2(c_1+c_3) q_{pw}$. Such linear analysis cannot capture the case of
nonlinear phase-twists, nor the case when the phase-diffusion
coefficient $D$ becomes negative (which happens outside the stable
band). As will be discussed below in more detail, the general
evolution of phase-twists is, to a large extend, governed by the
existence of the MAW and Homoclon coherent structures as illustrated
in Fig.~\ref{sketch}
\cite{maw,lutznew,clon,mm,zz,janiaud,kramer85,wound}.  {\em{(i)}} For
a wide range of parameters, both when the background wave is stable or
unstable, there exists a nonlinear coherent structure called a
``Homoclon' that corresponds to a particular local phase-twist
structure. This structure is linearly unstable, and acts as a
separatrix: ``smaller'' phase-windings decay, while ``larger'' ones
evolve to defects (arrows $A$ and $B$ in
Fig.~\ref{sketch}). {\em{(ii)}} There exist closely related but less
nonlinear structures referred to as Modulated Amplitude Waves
(MAWs). For small background wavenumber, the MAWs occur via the
forward Hopf bifurcation that occurs when a plane wave turns unstable;
in this case, the instability of the laminar background carries over
to the MAWs, and MAWs are often linearly unstable \cite{maw}.
Unstable MAWs often lead to a disordered state called phase-chaos, in
which patches of transient MAWs occur \cite{maw}; the structure of the
phase-gradient peaks in phase-chaos is comparable to those of MAWs
(arrows $B$ and $C$ in Fig.~\ref{sketch}).  For large background
wavenumber, MAWs can occur via a subcritical bifurcation, and may
become linear stable \cite{maw,lutznew,wound}. {\em{(iii)}} Homoclons
and MAWs are closely related and disappear in a saddle node
bifurcation for sufficiently large $c_1$ and $c_3$. After this has
happened, arbitrarily small phase-gradients diverge (arrow $D$ in
Fig.~\ref{sketch}) and defects form spontaneously
\cite{maw,lutznew,clon,mm}.

MAWs and Homoclons both are characterized by local concentrations of
phase gradient and corresponding dips of $|A|$. To complicate matters,
there is another coherent structure, known by the name Nozaki-Bekki
hole, which is also characterized by a dip of $A$, and which is a
source of waves with {\em different} wavenumber.  Before turning our
attention to the experimental relevance of these structures, a brief
overview of the coherent structures framework and the properties of
these three families of coherent structures is given. Readers familiar
with these structures can skip this introduction and go straight to
section \ref{secexp}.


\subsection{Coherent structures}

In this section I will briefly discuss the coherent structure
framework and list the most important properties of coherent MAWs,
Homoclons and NB Holes. Section \ref{secexp} focuses then on the
incoherent structures and their relevance for experiments.

The temporal evolution of coherent structures in the CGLE amounts to a
uniform propagation with velocity $v_{coh}$ and an overall phase
oscillation with frequency $\omega_{coh}$ \cite{saar1}:
\begin{equation}
A(x,t)= e^{- i \omega_{coh} t} a(\xi) e^{i \phi(\xi)},\hspace{5mm} \xi
:= x - v_{coh} t \label{coh_ans}
\end{equation}
When the ansatz (\ref{coh_ans}) is substituted into the CGLE, one
obtains a set of 3 coupled first order real ordinary differential
equations (ODE's) (see appendix \ref{app2}). These equations can be
written in a number of forms; the representation used here employs
$a$, the local wavenumber $q:=\partial_{\xi} \phi$, and $\kappa:=(1/a)
\partial_{\xi} a$ as dependent variables. Orbits of the ODE's
correspond to coherent structures of the CGLE.

The ODE's (\ref{ode1},\ref{ode2}) allow for a number of fixed points. For fixed
$v_{coh}$ and $\omega_{coh}$, there are two fixed points with $a\neq
0$, and these correspond to plane waves:
\begin{eqnarray}
q_{pw} &=& (\pm\sqrt{v_{coh}^2 + 4(\omega_{coh}+ c_3 )(c_1 +
      c_3)}+v_{coh})/2(c_1+c_3) \label{q_FP}\\ a &=& \sqrt{1-q_{pw}^2}
\label{a_FP}
      \\ \kappa & = & 0 \label{k_FP}
\end{eqnarray}
The relation between these two fixed points is that their
corresponding plane waves have the same frequency, $\omega_{coh}$, in
the {\em frame moving with velocity $v_{coh}$}. To see this, note that
plane waves in the coherent structures framework (\ref{coh_ans}) are
of the form: $\exp(-i \omega_{coh} t) a(\xi) \exp(i \phi (x-
v_{coh}t))$, where $\phi(\xi)=q_{pw} \times \xi$. Comparing this to a
plane wave $\propto \exp(i q_{pw} x- \omega_{pw}t)$ one finds that in
the stationary frame the frequencies of the plane waves given by
Eq. (\ref{q_FP})-(\ref{k_FP}) are: $\omega_{pw} = \omega_{coh}+ q_{pw}
v_{coh}$.  Demanding that the dispersion relation for plane waves
given by Eq. (\ref{qwave}) is satisfied yields: $\omega_{pw}=
\omega_{coh} + v_{coh} q_{pw} = - c_3 + q_{pw}^2 (c_1+c_3)$, from
which Eq. (\ref{q_FP}) immediately follows. The role of the parameter
$\omega_{coh}$ in the coherent structures ansatz can therefore be
interpreted as follows. If $\omega_{coh}$ would be $0$, than the phase
velocity of the plane waves would equal the propagation velocity of
the coherent structure. Due to the phase-symmetry of the CGLE, these
two velocities may (and usually will) be different, and in such case
$\omega_{coh} \neq 0$.

Below, three families of coherent structures are discussed: MAWs
(corresponding to limit-cycles), Homoclons (corresponding to
homoclinic orbits) and NB holes (corresponding to heteroclinic
orbits).

\subsubsection{Limit-cycles and MAWs}\label{limmaw}

\begin{figure}
  \vspace{-0cm} \epsfxsize=1.\hsize 
\mbox{\hspace*{-.01 \hsize}    \epsffile{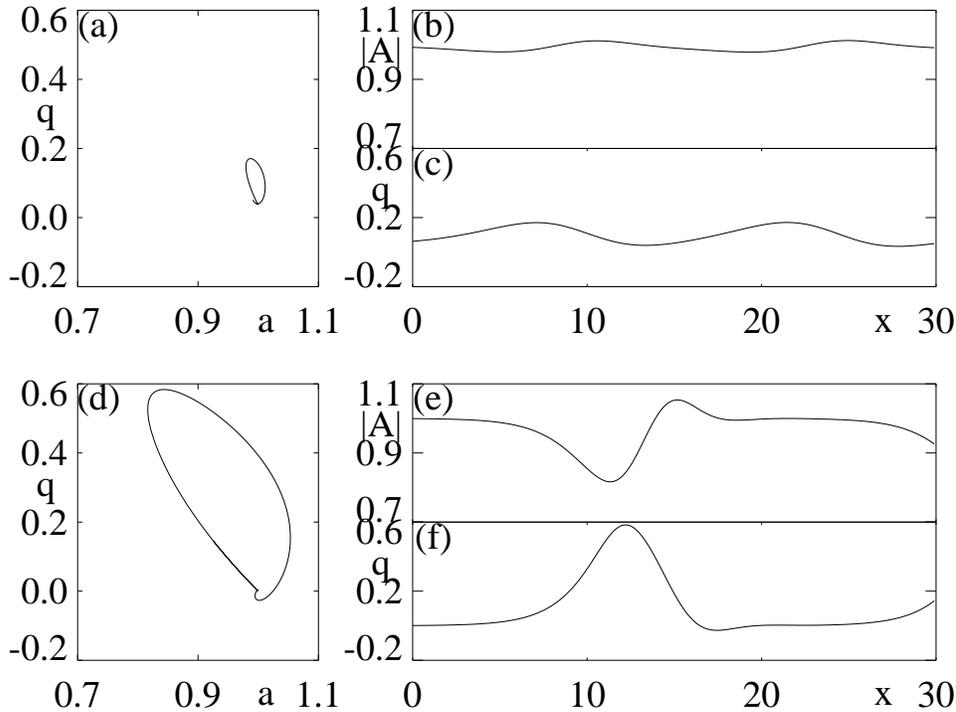} } 
\vspace{0.0cm} \caption{(a) For $c_1=c_3=1.1$, the ODE flow in $a,q$
space of a ``little'' limit cycle for $v_{coh}=0.381$ and
$\omega_{coh}=-1.11$. (b,c) Corresponding $|A|$ and $q$ profiles of
the coherent MAW characterized by a period $P\approx 15$ and
winding-number $\nu \approx 0.1$ (d) For $c_1=c_3=1.1$ one can also
obtain a ``large'' limit cycle for $v_{coh}=0.8864$ and
$\omega_{coh}=-1.1$.  (e,f) Corresponding $|A|$ and $q$ profiles of a
``periodic Homoclon'' with $P\approx 20, \nu \approx 0.13$. }
\label{figmaw}
\end{figure}

Limit cycles are periodic solutions to the ODE's
(\ref{ode1},\ref{ode2}). These orbits are structurally stable and
generically persist when the ODE's are perturbed. In particular, once
a limit cycle is obtained for certain values of $\omega_{coh}$ and
$v_{coh}$, limit cycles generically exist for nearby values of
$\omega_{coh}$ and $v_{coh}$. For many parameter regions, there are
two distinct limit-cycles that can be identified by their ``size'' in
phase space; for fixed $\omega_{coh}$, small(large) orbits occur for
small(large) values of $v_{coh}$ (see Fig.~\ref{figmaw}). The smallest
of these orbits occur when one of the plane wave fixed points of the
ODE's undergoes a Hopf bifurcation (and possibly a subsequent
drift-pitchfork bifurcations, see \cite{maw,lutznew}).  A weakly
nonlinear analysis shows that this bifurcation is supercritical
(forward) only when $c_1^2(1-6 c_3^2) + c_1 (2 c_3^3 + 16 c_3) - (8+
c_3^2) >0$; in many cases, this bifurcation is subcritical
\cite{janiaud,lutznew}. In the CGLE, these small cycles correspond to
periodically modulated plane waves that we refer to as MAWs. They can
be characterized by their spatial period $P$ and the ``average
winding-number'' $\nu$, which can be defined as $1/P \int_0^P dx
\partial_x \phi$ \cite{maw}. When following a particular branch of
solutions, $P$ and $\nu$ are functions of $c_1,c_3,v_{coh}$ and
$\omega_{coh}$ and can be obtained by a numerical analysis of the
ODE's \cite{maw,lutznew}.  (Fig.~\ref{figmaw}a).

The larger limit-cycles correspond to strongly nonlinear modulations
of plane waves in the CGLE (Fig.~\ref{figmaw}b). While there is a
whole family of these, again characterized by $P$ and $\nu$, the limit
where the period $P$ goes to infinity is most relevant\footnote{Since
Homoclons are unstable, finite $P$ solutions rapidly evolve to left
and right traveling holes, which are well-approximated by interacting,
infinite $P$ solutions.}; this particular family of CGLE solutions are
referred to as Homoclons (see below). The two families of limit cycles
merge and disappear in a saddle-node bifurcation; for details see
\cite{maw,lutznew}.

Since there is no closed expression for the full MAW family available,
a detailed analysis of their range of existence and stability
necessarily involves numerical calculations. Even when one limits
oneself to the ``small, weakly nonlinear'' branch, one still would
like to obtain the existence and spectrum as a function of $c_1,c_3,P$
and $\nu$. Here I will summarize the main results; more details can be
found in \cite{maw,lutznew}.

\begin{itemize}
\item For fixed $P$ and $\nu$, there is a certain range in $c_1,c_3$
space where the MAWs do exist. For large $c_1$ and $c_3$ this range is
limited by the aforementioned saddle-node bifurcation, while for small
$c_1$ and $c_3$ it is essentially the Hopf/pitchfork bifurcation that
limits their existence.
\item For small $\nu$, all MAWs are linearly unstable. There are
roughly two instability mechanisms that compete. For small $P$, an
``interaction'' mode is dominant. This mode acts as to break the
periodicity of the MAWs, but the number of phase twists stays the same
\cite{maw}. For large $P$, the ``splitting'' instability is
dominant. This latter basically is the Eckhaus instability acting on
the long patches of plane wave that occur for large $P$; this
instability tends to generate new phase inhomogeneities between the
peaks of the MAWs, roughly doubling the number of phasetwists.  The
dynamical state that occurs for small $\nu$ due to the competition of
these instabilities is called phase chaos, which, for $c_1$ and $c_3$
close to the so-called Benjamin-Feir-Newell curve ($c_1 c_3 = 1$) can
be approximated by the Kuramoto-Sivashinsky equation
\cite{maw,review,ch}. I am not aware of any experimental observation
of phase chaos in one space dimension.
\item For larger values of $\nu$, MAWs may become stable and phase
chaos can be suppressed. Such stable structures have also been called
``compression waves'' or ``wound up phase-chaos''
\cite{maw,lutznew,janiaud,wound,chif}.
\end{itemize}

\subsubsection{Homoclinic orbits and Homoclons}

Homoclinic orbits start from one of the plane wave fixed points, make
an excursion through phase space and then return to the same fixed
point. They are structurally stable and occur as a one-parameter
family. Once such an orbit is obtained for certain values of $v_{coh}$
and $\omega_{coh}$, when $\omega_{coh}$ is varies, $v_{coh}$ has to be
varied accordingly to obtain a homoclinic orbit again. These
homoclinic orbits can be obtained by letting the period $P$ of the
limit cycles diverge. Similar to these limit cycles, there are two
branches of homoclinic solutions; the small ones will be referred to as
$P \rightarrow \infty$ MAWs, while the ``large, strongly nonlinear''
of these two correspond to Homoclons\footnote{The strongly nonlinear
but periodic structures do not play an important role and hence do not
have their own name; they can be referred to as ``periodic
Homoclons'', although this is not very elegant.}. These are
propagating structures that connect two plane waves of equal
wavenumber in the CGLE. Their core region is characterized by a
concentration of local wavenumber $q$ (or equivalently, a twist of the
phase $\phi$), and a dip of the amplitude $|A|$ \cite{clon,mm}. They
can be seen as local phase twists that glide through a background
plane wave; often their total phase twist is of the order of, but not
precisely equal to, $\pi$. Their winding number $\nu$ simply
corresponds to the wavenumber of the asymptotic plane wave.

For fixed $c_1,c_3$ and $\nu$, there are left and right moving
homoclons, related by left-right symmetry of the CGLE (which maps $q
\leftrightarrow -q$). Their propagation velocity is typically much
larger than the nonlinear group-velocity of the plane wave they glide
through; they are neither sources nor sinks. In the comoving frame,
they experience an incoming wave ahead and leave an outcoming wave
behind.

For positive $c_1$ and $c_3$, solutions occur for positive (negative)
$v_{coh}$ when the $q$-profile peak is positive (negative). When $c_1$
and $c_3$ are both negative it follows (by complex conjugation of the
CGLE) that the signs of velocity and $q$-profile peak are opposite.

As for the MAWs, no analytic expression exists for the Homoclons, and
a numerical analysis has revealed the following key points:
\begin{itemize}
\item The range of existence of Homoclons is limited by the same
saddle-node bifurcation as the MAWs. Once a Homoclon has been obtained
for certain values of $c_1$ and $c_3$, this solution persists when the
coefficients are decreased towards zero; in fact these
Homoclon-solutions smoothly connect to the analytically known
saddle-point solutions of the real ($c_1=c_3=0$) Ginzburg-Landau
equation \cite{kramer85}. For large values of $c_1$ and $c_3$, no
Homoclons exist, and generic initial conditions form defects; the CGLE
is then in the defect chaotic regime \cite{maw}.
\item The spectrum of the Homoclons consists of two parts, a
continuous part associated with the asymptotic plane waves, and the
discrete part associated with a few core modes \cite{clon}. For all
cases that I am aware of, there is one single unstable core mode; this
unstable eigenmode is associated with the aforementioned saddle-node
bifurcation. 
\end{itemize}

With hind-sight, the existence of Homoclons and their main properties
are not entirely surprising. Suppose, for some values of $c_1$ and
$c_3$, that one can construct two ``generic'' initial conditions, one
that does not evolve to defects, another that does form a defect. When
$c_1 c_3<1$ this is certainly possible: the state $A=1$ is stable and
nearby initial conditions will not form a defect, while a state where
$A=1$ except for an interval where $A\sim \exp(i Q x)$ will form
defects when $Q$ is sufficiently large. Now imagine that one
interpolates between these two initial conditions; let's call the
interpolation parameter $b$. In the simplest case there will be a
single transition value for $b$ where the dynamics changes from
non-defect to defect forming. Clearly, at the transition point
something special must happen, and the simplest scenario here is that
the time it takes for a defect to form diverges there. This may happen
in a variety of ways, but the simplest case would be the formation of
a saddle-point like structure: the Homoclon.  Since one needs only one
parameter to vary, it is reasonable to expect a single unstable
eigenmode.

Note that in the defect chaos regime this scenario breaks down, since
only non-generic initial conditions (perfect plane waves) will not
form defects: indeed, no Homoclons exist here.

This scenario can be put on firmer ground for the real Ginzburg-Landau
equation, for which analytical results are available
\cite{kramer85}. The dynamics of this equation is governed by a
Lyapunov functional that is decreasing over the course of time. Of
course, one can create saddle points that correspond to unstable
coherent structures. The simplest example of this is the standing hole
solution $A=\tanh(\sqrt{1/2} x)$, which indeed corresponds to a
homoclinic orbit of the ODE's (\ref{ode1},\ref{ode2}).  It is a
numerically easy task to trace a coherent Homoclon orbit as a function
of $c_1$ and $c_3$, and one finds that in the $c_1=c_3=0$ limit, a
$\nu=0$ Homoclon smoothly deforms to this saddle-point solution of the
real Ginzburg-Landau equation; the propagation velocity of the
homoclons thus goes to zero in the relaxational limit.

\subsubsection{Heteroclinic orbits and Nozaki-Bekki Holes}
The heteroclinic orbits that are relevant here start at one plane-wave
fixed point and end up on another plane-wave fixed point.  They can be
obtained in a closed analytical form \cite{nb,saar1} and the
corresponding CGLE structures are the so-called Nozaki-Bekki (NB)
holes which connect waves of {\em different} wavenumbers $q_1$ and
$q_2$.

For the unperturbed CGLE, once such an orbit is found for certain
values of $v_{coh}$ and $\omega_{coh}$, there will generically be
heteroclinic orbits for nearby values of $v_{coh}$ and
$\omega_{coh}$. As pointed out in \cite{saar1}, this is not the
generic situation one expects (on the basis of counting arguments) for
these orbits. Indeed it was shown that small perturbations of the CGLE
destroy most of the NB-orbits/holes \cite{nbunst}, and one is left
then with a discrete family of solutions: almost all NB holes are
structurally unstable solutions to the 1D CGLE.  The dynamical states
that occur in this situation are discussed in \cite{nbunst}.

Once $q_1$ and $q_2$, the wavenumbers of the adjacent plane waves, are
fixed, $v_{coh}$ and $\omega_{coh}$ can be obtained from a simple
phase matching argument, which states that in the comoving frame the
apparent frequency of the two waves needs to be equal to
$\omega_{coh}$.  Some manipulation yields that $v_{NB} =
(q_1+q_2)(c_1+c_3)$ and $\omega_{NB} = - q_1q_2(c_1+c_3) -
c_3$. Comparing this propagation velocity to the nonlinear group
velocities of the adjacent plane waves, $2 q_1 (c_1+c_3)$ and $2 q_2
(c_1+c_3)$ respectively, it immediately is clear that the propagation
velocity of the NB holes is in between these group-velocities. A more
thorough analysis \cite{saar1} shows that NB holes are always {\em
sources}, i.e., in the comoving frame they send out waves.

Since the NB holes are known in closed form, a number of results for
the range of existence and stability are available. In particular, the
range of stability of NB holes is limited by the instabilities of
their adjacent plane waves on one side, and a core instability on the
other side.  This yields a band of stable NB holes, which is mainly
located in the regime where $c_1$ and $c_3$ have opposite signs
\cite{review}.

\section{Incoherent dynamics and experiments}\label{secexp}

When one is in the lucky position that an experiment shows interesting
spacetime dynamics of local structures, some information on how to
identify the structures that occur can be found in section
\ref{subiden}. But if the system by itself just produces simple plane
waves, which often seems to be the case, some manipulations need to be
done. The good news is, that for all coefficients of the CGLE it is
possible to generate transient Homoclons by locally perturbing the
system, and when the wavenumber of the plane states can be controlled,
also a number of MAW states can be generated. 

In most cases one would expect to observe incoherent rather than
coherent MAWs, Homoclons and NB-holes. From the wide range of behavior
that occurs in the CGLE, I will highlight a number of states that
hopefully will be accessible in experiments. The examples given below
are intended to inspire particular experiments, and are not fully
worked out recipes. Before going on, let me briefly discuss some of
the properties of candidate systems.

\begin{itemize}
\item{\bf Boundary conditions.} Boundary conditions will play an
important role. Some systems, such as Rayleigh-B\'enard
wall-convection in rotating cells \cite{liu,53lutz} or sidewall convection
in annular containers \cite{chif} have periodic boundary
conditions. In this case the linear group-velocity term of the CGLE
(see appendix \ref{app}) can be removed by going to the comoving
frame. The main finite size effects to be expected are then the
discretization of possible wavenumbers and the fact that a source will
necessarily be accompanied by a sink.  The left and right boundaries
of long rectangular systems, such as heated wire convection cells
\cite{hw,wvdw} and sidewall convection \cite{burg,garnier}, may have a
more severe effect on the dynamics. First of all, the group-velocity
can no longer be ignored here.  In addition, in some systems the
boundaries may act as sources that send out waves of a selected
wavenumber \cite{mysource,source}. Of course this can be used in experiments,
in particular when the selected wave can be made linearly unstable.

\item{\bf Control-parameters.} For all systems, the dimensionless
distance to threshold, $\varepsilon$ (see appendix \ref{app}) is
experimentally accessible. While $\varepsilon$ may be scaled out of
the equation for the amplitude $A$, it does effect the spatial and
temporal scales. For example, a sudden change in $\varepsilon$ has the
effect of virtually ``stretching'' or ``compressing'' the spatial
scale of $A$. This could be used to manipulate the {\em effective}
wavenumber and the period $P$ of MAWs (see below).

When the group-velocity cannot be ignored due to boundary effects,
changes in $\varepsilon$ can change the instability of plane waves
from convective to absolute. The amplitude equations no longer scale
uniformly with $\varepsilon$ in this case, and the stability of
sources pinned at boundaries may change (for more on this in the
context of left and right traveling waves see \cite{mysource,source}).

In some cases, other experimental parameters are accessible, like the
depth of the heated wire below the surface \cite{hw,wvdw}. Such
changes will affect the coefficients of the CGLE, although often these
are not known, let alone their dependence on experimental parameters.

\item{\bf Local manipulation.} Usually the initial conditions cannot
be chosen at will, in stark contrast to what is done in numerical
studies. Nevertheless, some manipulations of dynamical states are
usually possible. In convection systems, a short burst of local
heating can be used to perturb the system. Such perturbations can be
used to generate fronts (see appendix \ref{app}) or possibly to
generate Homoclons (see below). One may also imagine that by periodic
local heating one can manipulate wavenumbers or the period of MAWs.
In systems with a free fluid surface, mechanical manipulations such as
briefly touching the surface or depositing a few drops of oil
\cite{chif} may also be used as (crude) ways of perturbing the system
locally.

\item{\bf CGLE Coefficients}. For most systems, the coefficients of
the CGLE have not been calculated, making a detailed comparison to
theory rather cumbersome. By careful experiments (see appendix
\ref{app}) one should be able to get good estimates for most of their
values \cite{burg,liu,wvdw,kroket}.

\end{itemize}

\subsection{Incoherent MAW dynamics}

A comprehensive overview of MAW properties can be found in
\cite{maw,lutznew} and I will mention the most relevant properties in
the list of suggestions for experiments given below.

\subsubsection{Stable and unstable MAWs}

When plane waves turn unstable against the Eckhaus instability, MAWs
occur.  When the wavenumber of the initial plane wave is sufficiently
large (i.e., for $c_1$ and $c_3$ far away from the BFN curve), the
bifurcation to MAWs is subcritical and such MAWs may be linearly
stable \cite{lutznew,janiaud,wound}. For such structures one can
obtain their velocity, period and winding-number (provided the CGLE
coefficients are known), and confront these with numerical
calculations of MAW profiles. In the subcritical regime, it may be
possible in this regime to excite MAWs by sufficiently strong local,
periodic forcing of the waves. Such exciting of MAWs can of course
also be done at the boundary of a system, although periodic boundaries
are probably best suited for such experiments.

\begin{figure}
  \vspace{-0cm} \epsfxsize=1.\hsize 
\mbox{\hspace*{-.01 \hsize}    \epsffile{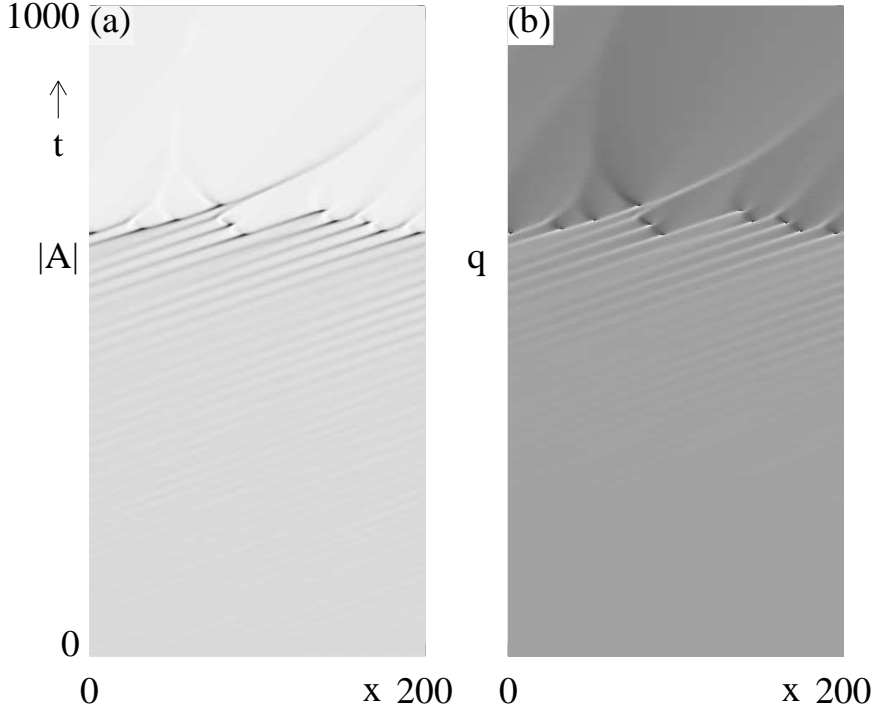} } 
\vspace{0.0cm} \caption{Space-time plot of the evolution of the
Benjamin Feir instability of a plane wave with initial wavenumber
$q_{pw}=0.41$, for $c_1=c_3=0.8$ in a system with periodic boundaries
and size 400; only part of the system is shown. The slow growth of a
transient MAW can be observed, and since this MAW is ``beyond the
saddle-node'', it does not saturate but develops defects. This
dynamics is the one depicted by arrow D in Fig.~1, and is very
reminiscent of defect formation observed in hydrothermal waves (see
Fig.~6 of \cite{chif}) and rotating Rayleigh-B\'enard convection
\cite{liu}.}
\label{figtrans}
\end{figure}

Starting from stable MAWs, a number of MAW properties may be probed.

\begin{enumerate}
\item When the period of MAWs is altered (by periodic local heating or
sudden changes of $\varepsilon$) and increased, two possible types of
dynamics may occur. When $c_1$ and $c_3$ are relatively small, the MAW
may become prone to the splitting instability \cite{maw,lutznew}, and
new peaks in $q$ may be generated between the old ones, effectively
decreasing the period $P$. When $c_1$ and $c_3$ are larger, the effect
of increasing $P$ may be to cross the saddle-node bifurcation, and
defects will then occur \cite{maw}. Since the occurrence of this
saddle-node appears to be a crucial ingredient for the understanding
of the transition from phase to defect chaos, this would be extremely
interesting to see.
\item When the period of MAWs is decreased, the interaction
instability may kick in, which breaks the periodicity of the MAWs
\cite{maw}.
\item When the winding number of MAWs becomes sufficiently small, MAWs
become unstable (see \cite{lutznew,wound}). In principle, the
effective wavenumber changes when $\varepsilon$ is varied, but at the
same time the effective period $P$ changes also. Therefore, a change
of $\varepsilon$ can either lead to MAW instabilities or
defect-formation, depending on which instability is closest. If it is
possible to manipulate $P$ by local periodic heating, then changing
this period according to the changes made in $\varepsilon$ may be a
way to change the effective value of $\nu$.
\item While it is difficult to say anything general about the MAW
profiles, it is known that the minimal value of $|A|$ and the extremum
of $|q|$ both grow as a function of the period. In fact, near the SN
they decrease ($|A|$) and increase ($|q|$) sharply, which could be
used to estimate if this bifurcation is close by.

\end{enumerate}

\subsubsection{Transient MAWs.}

When the plane wave is stable but close to its  Eckhaus
instability, perturbations of this plane wave will decay rather
slowly, and transient MAWs can be formed. By measuring the decay time
as a function of the wavenumber, one can make an estimate for the
proximity of the Eckhaus instability (see \cite{chif}).

When a plane wave becomes unstable due to the Eckhaus instability, the
first stages are characterized by the development of periodic
modulations. This period is determined by the most unstable wavenumber
of the Eckhaus instability. When these modulations do not grow out to
phase chaos or MAWs, but instead grow without bound, defects will
eventually occur; a typical example is shown in
Fig.~\ref{figtrans}. The transient MAWs are, as usual, characterized
by their period $P$ and winding-number $\nu$ (which is equal to the
wavenumber of the initial wave). Defect formation is then expected to
occur when, for the current values of $c_1$ and $c_3$, $P$ and $\nu$
lie beyond the appropriate saddle-node bifurcation
\cite{maw,lutznew}. When $P$ and $\nu$ are relatively close to the
saddle-node, the formation of defects can be rather slow, and
transient holes can be observed; these are clearly incoherent (see
\cite{chif,53lutz} for possible experimental realizations of this).

\subsection{Incoherent Homoclon dynamics}

Homoclons are never linearly stable; over the course of time, they
either slowly decay or grow out to form a phase-slip
\cite{clon,mm}. It will be assumed that their dynamics is studied in a
background of linearly stable plane waves. In earlier studies it was
found that the instability of the Homoclons is due to a single
``core''-mode, and its unstable eigenvalue is the one which changes
sign at the saddle node bifurcation where MAWs and Homoclons merge and
disappear \cite{maw,lutznew}.

This single weakly unstable eigenmode is reflected in the dynamics of
the Homoclons.  Many sufficiently localized wavenumber-blobs will be
attracted to the 1D unstable manifold of the Homoclon; subsequently
they then evolve along this manifold, in either the ``decay'' or the
``phase-slip'' direction.  One can loosely think of the homoclinic
holes as unstable equilibria, or separatrices, between plane waves and
phase-slips \cite{clon,mm}.

\begin{figure}
  \vspace{-0cm} \epsfxsize=1.\hsize 
\mbox{\hspace*{-.01 \hsize}    \epsffile{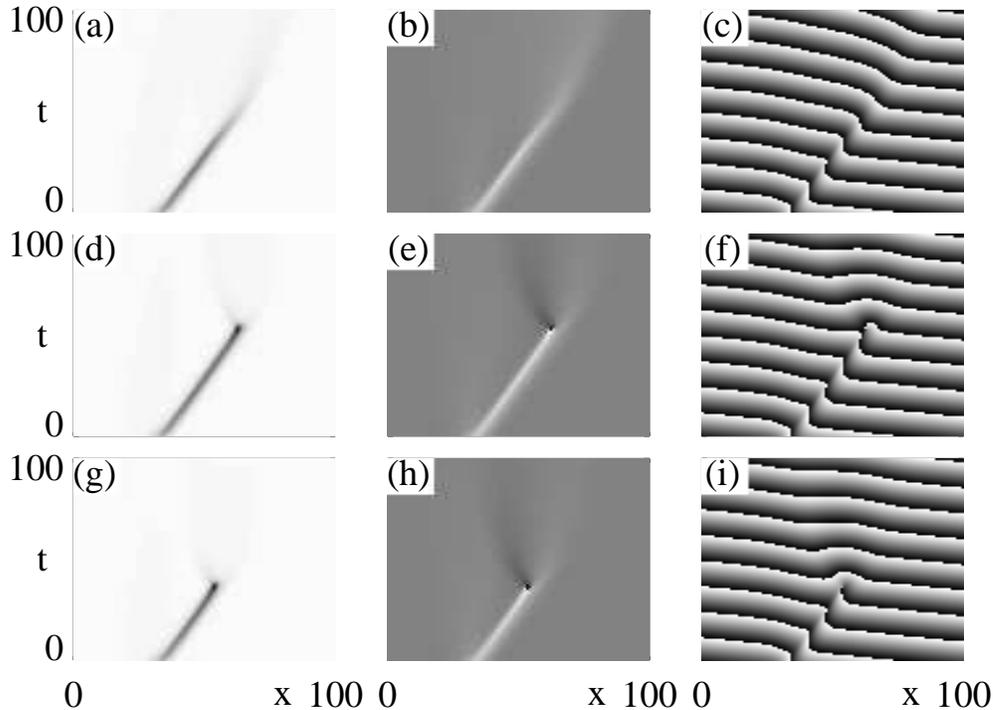} } 
\vspace{0.0cm} \caption{Incoherent Homoclons for $c_1=c_3=0.5$.  The
first column shows the evolution of $|A|$ (white: $|A|=1$) while the
second column shows the evolution of $q$ (grey: $q=0$, light: $q>0$).
The third column shows the evolution of the phase of an artificially
created ``fast'' field: for that $A$ has been multiplied with a plane
wave solution.  (a,b,c) Initial phase-twist just below Homoclon
equilibrium.  (d,e,f) Initial phase-twist just above Homoclon
equilibrium.  (g,h,j) Initial phase-twist further above Homoclon
equilibrium; the only difference to d,e,f is the shorter lifetime of
the incoherent Homoclon here due to a different initial condition.}
\label{fig3clon}
\end{figure}

These properties can be probed as follows: Suppose one has a stable
plane wave, and locally perturbs it by generating a twist in the phase
field of $A$. When this twist is small, the linear stability of the
plane waves will govern the dynamics and such a perturbation will
decay diffusively (downward pointing arrow A and B of
Fig.~1). However, when it is strong enough, and the initial conditions
``passes through the Homoclon separatrix'', the twist will grow out to
form a defect (upward pointing arrows A of Fig.~1). When tuning the
initial twist so as to be in between the decay and defect scenario,
one can obtain a Homoclon that exists for an arbitrarily long time (In
practice, noise may limit this time.). The non-trivial prediction is
that, since there is only one unstable eigenmode, one only needs a
one-parameter family of initial conditions to ``hit'' the unstable
Homoclon solution. This situation was explored already in \cite{clon}
in the intermittent regime, but it is also possible to generate
homoclons in the ``laminar'' CGLE regime dominated by stable plane
waves (see Fig.~\ref{fig3clon}). In addition, the profiles of
incoherent homoclons evolve, within good approximation along a
one-dimensionally family; if one takes many snapshots of incoherent
homoclons and orders them by their extremal phase-gradient $q_{ex}$,
profiles with the same $q_{ex}$ should look fairly similar (see
\cite{clon} for a theoretical check of this). These properties open up
the possibility for a number of experiments that follow below.

\subsubsection{Creating transient Homoclons}

As discussed above, transient Homoclons can be generated in a large
range of coefficient space, provided one can locally perturb the
system sufficiently. A possible experimental protocol would be as
follows.
\begin{itemize}
\item Generate a stable plane wave, and find a way to locally perturb
this wave. As discussed above, thermal systems could be perturbed by
local heating and many systems could be perturbed by mechanical means.
Whatever method one uses, one would expect to be able to tune the
strength of the perturbation.
\item Clearly, for small enough perturbations no defects will be
formed, but instead the perturbation will drift with the group
velocity and decay diffusively. The question is: can one make a large
enough perturbation that grows out to a defect? This is the necessary
ingredient; and this is where clever experiments will be the best way
to find out what perturbations are most successful. From a theoretical
point of view, it is known that perturbations that ``twist'' the
phase-field are most effective. I would imagine that the effect of
local heating is a local increase of the effective value of
$\varepsilon$, and the resulting change in time and spatial scales
will certainly lead to the generation of phase-twists. If these are
done in the middle of the system, likely two pairs of twists are
generated (at either side of the perturbed region), so it may be
advantageous to perform the perturbations at the edge.
\item Denote the control parameter that controls the strength of the
perturbation as $V$. If, defects are formed for $V=V_2$, while for
$V=V_1$ the perturbation decays, the experimental protocol is then to
repeat this experiment for a number of values between $V_1$ and $V_2$
to obtain a critical value $V_c$ between decay and defect
generation. Approaching this value, the theory predicts that transient
Homoclons of longer and longer lifetimes occur, and that their
lifetime diverges as $-\lambda^{-1} ln (|V-V_c|)$ \cite{mm}, where
$\lambda$ is the unstable eigenvalue of the Homoclon \cite{mm}.
\item The theory predicts that the tuning of a single parameter is
sufficient to come arbitrarily close to a coherent Homoclon, i.e., if
one has two control parameters, say $V$ and $W$, than there should be
a continuous curve in $V,W$ space where the Homoclon lifetime
diverges.
\item Finally, if one is able to tune the wavenumber of the plane wave
that the Homoclon propagates into, this also can be used as a
parameter. This property was explored in \cite{mm} in the regime where
the coefficients $c_1$ and $c_3$ are such that the defects generated
by Homoclons lead to the birth of new homoclons and periodic
hole-defect states occur. Even when this is not the case, i.e., in the
``laminar'' regime, the lifetime of a Homoclon can still be controlled
by the asymptotic wavenumber.
\end{itemize}

\subsubsection{Spontaneous incoherent Homoclon generation}

There are a number of states in which incoherent Homoclons are
generated. First there is the hole-defect chaos that occurs in a part
of the so-called spatial-temporal intermittent regime of the CGLE
\cite{clon,mm,chat2}; as far as I am aware, this state has not yet
been experimentally observed. In experiments it is often possible to
have sources between left and right traveling waves, and in that case
a sufficient lowering of $\varepsilon$ leads to the creation of
unstable sources that send out irregular phase-twists that can lead to
transient Homoclons \cite{wvdw}.

In both of these states one may study the $a$ and $q$ profiles of
snapshots of the holes, and see if they indeed can be ordered as a one
parameter family as is the case for the CGLE \cite{clon}. In addition,
when following $q_{ex}$ as a function of time for a hole, the time
derivative $\dot{q}_{ex}$ should be a function of $q_{ex}$ only
\cite{mm}; this should be revealed in scatter-plots of $\dot{q}_{ex}$
versus $q_{ex}$. Finally it should be remarked that the late stages of
defect formation in hole-defect dynamics also is expected to show some
universal features; for more see \cite{mm}.

\subsection{Nozaki-Bekki holes}

I will not spend much time on the incoherent dynamics of Nozaki-Bekki
holes since there is already a very large body of literature available
\cite{nbunst,nbint,review}. There is a potentially large range of
dynamical behaviors possible for NB holes, due to either dynamical or
structural instabilities.  When one would observe dynamical NB holes
in an experiment, it may be interesting to measure the time-scales as
a function of $\varepsilon$.  For a CGLE-like structure, these should
scale as $\varepsilon^{-1}$, but if the dynamics is due to the
structural instability one may expect a larger exponent here, since
the deviation from the CGLE can also be expected to scale with
$\varepsilon$; on the other hand, when NB holes are dynamically
unstable, the structural instability may be irrelevant.

\subsection{Identification of local structures}\label{subiden}

What sort of states can be expected to be seen in experiments? In the
light of the experimental data available and our discussion of the
properties of the various coherent structures, three possibilities
seem most common: {\em{(i)}}: Stable MAWs (either in small systems or
for high winding-number) \cite{janiaud,chif,bot2} {\em{(ii)}}:
Transient MAWs that evolve to defects \cite{chif,liu} {\em{(iii)}}
Holes that may be Homoclons or NB Holes.  Since cases {\em{(i)}} and
{\em{(ii)}} have been discussed already in the sections above, only
the distinction between Homoclons and NB holes will be addressed.

When one observes coherent holes, this situation appears simple: When
the two wavenumbers of the adjacent waves are equal, they may be
Homoclons, when the are different, they may be NB holes. The problem
is that in realistic situations, both wavenumbers always will be
different, due to noise etc; as can be expected due to the structural
and dynamical instabilities of these structures, typically incoherent
holes are observed. For incoherent structures, one cannot take a
single snapshot and compare this to the profile of either Homoclons or
NB holes. Also, the wavenumbers at both sides of an incoherent
Homoclon are, due to the combination of an evolving core and phase
conservation, {\em not} equal (see \cite{clon,mm}).

There is, however, a clear difference between incoherent Homoclons and
NB holes, which manifests itself in the direction of the full
nonlinear group-velocity of the adjacent waves. I would even like to
go so far as to propose this as a definition: \newline {\em In the
frame moving with the incoherent hole, incoherent NB holes send out
waves, while incoherent Homoclons have one incoming and one outgoing
wave}. \newline Needless to say, checks on the direction of the
group-velocity are also useful for {\em coherent} holes.

An important consequence of this is that two incoherent NB holes
cannot sit immediately next to each other, but need to be separated by
a so-called {\em sink} \cite{saar1} (unless they have very different
propagation velocities). This feature can be used when it is difficult
to measure the group velocity, or when this velocity becomes close to
the propagation velocities of the holes (which is the case for small
$\varepsilon$, where the group velocity becomes close to the linear
group velocity $s_d$; see appendix \ref{app}).

I will now briefly discuss some earlier experimental work in which
holes are observed; for a more general overview of earlier
experimental work, the reader is referred to \cite{review,ch}.  In
\cite{lega}, convection of a low Prandtl number fluid in an annular
container is studied. Heated wires embedded in the walls of this
container yield horizontal temperature gradients which result in a
stable two-concentric-roll pattern. These roles become unstable to an
oscillatory instability when an additional vertical temperature
gradient is imposed, leading to one-dimensional waves that travel
along the annulus. For Rayleigh numbers larger than 1.2 $Ra_{crit}$,
these waves are stable, and the authors focus on the case that $Ra=
1.1 Ra_{crit}$, when the homogeneous wave becomes unstable and strong
but slowly varying amplitude and wavenumber modulations develop along
the pattern. The authors conclude that these modulations presumably
are not described by (coupled) CGLEs; the modulation does, however,
yield a local area of relative large wavenumber in which holes are
generated.

These holes propagate for a finite amount of time, and in many cases
lead to the formation of defects. To quote the authors: ``We observe
the spontaneous creation of traveling dips in a traveling wave
pattern. They appear in a region of low amplitude and propagate
through it. {\em They change their depth and the value of the phase
jump at their core as they move}''. While it is difficult to make
definitive statements, it may very well be that the holes are formed
in a manner similar to that shown in Fig.~3, i.e., due to the
instability of a wave; the authors in fact claim this in their
conclusion. At the point of writing of this paper, only NB holes were
known and the authors speculate that the structures they observe are
NB holes, but I doubt that: now that we know that beside NB holes,
there are also Homoclons and ``ghost'' states such as shown in Fig.~3,
the question what sort of holes where observed in this experiment is,
in my opinion, open.

In \cite{fleselles}, oscillatory Rayleigh-B\'enard convection in a low
Prandtl number fluid (argon gas) is studied. The state of 'coupled
oscillators' that the authors observe may become chaotic, in which
cases holes occur (see their Fig.~6). The holes are ``characterized by
a strong amplitude dip and a phase jump localized at their cores.''
They have a finite lifetime and terminate in defects. ``The slower
they move, the deeper the amplitude dip''.  The incoherent motion of
the holes, their evolution to defects and the fact that a defect can
generate a pair of holes with opposite phasejump (see Fig.~6 of
\cite{fleselles}) all are strongly reminiscent of the dynamics of
incoherent Homoclons. As before, when this paper was written, only NB
holes were known and so naturally the authors speculate that their
holes are of this form, but it seems fair to say that the situation is
unclear; in fact, the dynamics shown in their Fig.~6 could at least as
well be due to homoclinic type holes, also because strong evidence for
significantly different wavenumbers at both sides of the holes is
absent \cite{privcom}.

More recently, a careful study of holes occurring in hydrothermal
waves in a laterally heated fluid layer was carried out. The system
consisted of a rectangular container of length 25cm and width 2cm, in
which left and right traveling of typical wavelegnth 5mm occured. The
coefficients $v_{gr}$, $\xi_0$ and $\tau_0$ where all measured, and
$\varepsilon$ was at a value of 0.19. The dynamical state that occurs
then (see Fig.~2 of \cite{burg}) is that of an unstable source which
separates left and right traveling waves; this source sends out
perturbations that develop into (incoherent) holes. The authors claim
that these holes are of NB type. By measuring the profile of such hole
(in particular their left and right wavenumber and the value of the
amplitude minimum), and comparing these to values know from the NB
family, they estimate that $c_1 \approx -1.5$ and $c_3 \approx 0.4$;
the authors claim that similar values of $c_1$ and $c_3$ were obtained
for a number of different holes send out by the unstable source.
Unfortunately, no independent measurements of the coefficients $c_1$
and $c_3$ could be extracted for the hydrothermal waves.  Using the
values of the coefficients quoted by the authors, I studied the
behavior of the (coupled) CGLEs here, using the values of the
coefficients quoted by the authors (including a rescaled
group-velocity of 1.7), and found indeed an unstable source that sends
out hole-like perturbations (see \cite{mysource} for an explanation of
the instability of the source). In the simulations these holes are
quite incoherent, and it is difficult to see whether they are
homoclinic or heteroclinic.

I have no acces to the raw data, and the evidence for NB holes
presented is impressive. Nevertheless, there are some puzzling aspects
to this experiment. For the values of coefficients and wavenumbers
quoted, NB-holes are quite unstable; I found in simulations of a
single CGLE that even the generation of transient unstable NB holes is
difficult to achieve for generic initial conditions; I do not know of
any mechanism that would generate unstable NB holes, and it is
therefore surprising, but not impossible, that these holes are seen in
experiment! Another thing that worries me is the size of the holes,
and their ill separation.  The snapshot of a NB hole that the authors
show in their Fig.~4 has an extension of roughly 50mm on the scale of
their spacetime diagram Fig.~2\footnote{since $\xi_0
/\sqrt{\varepsilon}=2.5$ mm \cite{burg}.}, and so I would have
expected to see clear differences in wavenumbers and sinks in the
spacetime-diagram. But this diagram shows the occurrence of holes that
are close together (distance approximately 10-20 mm), without any
indication of a sink in between that would separate them if they would
be NB holes. Also, NB-holes are sources, so they should send out
waves; again, I don't see any evidence for this.

An alternative, but possibly less satisfying interpretation of the
data goes as follows. First note that the total time shown in the
spacetime plot is quite short also (105 sec, while $\tau_0
/\varepsilon = 8$ sec \cite{burg}), and that all propagation is
completely dominated by the linear group velocity. The holes, when
they are just send out from the source, form defects, and the
pre-defect holes appear reminiscent of (transient) Homoclons; their
dynamics is quite fast and so the wavenumbers to the left and right
should be substantially different; in fact they seem so incoherent,
that detailed comparison to either NB or homoclinic holes seems fairly
hopeless. As far as I understand, the holes studied by the authors are
the ones formed after these defects have occurred. If one now takes
the single CGLE and studies defect formation for the coefficients
quoted by the authors, one finds (for example by taking an initial
wave of high wavenumber ($0.6$)), that after the last defect was
formed, a transient state persisted for a few timeunits that is
characterzized by a dip of $|A|$, and two wavenumber ``blobs'' (that
come from the decaying defect); one could interpret these as patches
of waves with different wavenumbers to the left and right of this dip.
When one switches on the group velocity, such structures are swept
away, and could look similar to traveling holes.  These structures
decays than fairly rapidly (of the other of 10-20 time units in CGLE,
which would be 80-160sec in the experiment), but are, in my opinion,
not related to any coherent or incoherent hole state. Note that the
occurrence of short timescales is inherent in small-$\varepsilon$
experiments in finite systems; either very large systems or periodic
boundary conditions may make more detailed comparison of the holes
send out by unstable sources in hydrothermal waves possible in the
future, and I'm looking forward to see what would come out of such
experiments!

Finally, in \cite{bot}, the dynamics of holes and the formation of
defects is studied in the Taylor-Dean system. In this system a
traveling wave with negative phase diffusion occurs, and phase
gradients concentrate and lead to defects (in related work, these
authors observed stable MAWs \cite{bot2}). The dynamics depicted in
Fig.~10 and 11 of \cite{bot} suggest that transient Homoclon-like
structures play a role here. Finally, in a recent study of sources and
holes in a heated wire convection experiment \cite{wvdw}, the dynamics
of holes send out by unstable sources is studied. The temporal
evolution of the profile of the holes shows properties reminiscent of
that of Homoclons; nevertheless some questions remain open.

Notwithstanding all these indications, in my opinion neither NB holes
or Homoclons have been observed unambiguously yet. Experiments in which
the coefficients of the amplitude equation are measured independently
and compared to the profile and evolution of holes, or precise
comparison of group-velocities and propagation velocities of holes may
well show (unstable) NB holes. Similarly, experiments in which initial
phasetwists can be shown to lead to either long or short lived holes
(similar to Fig.~4) may yield more definite experimental evidence for
Homoclons.

\section{Conclusion and Outlook}\label{secout}

In this paper I hope to have given a broad and easy accessible
introduction to the dynamics of MAWs and Homoclons, and some
clarification on the Homoclon/NB hole issue. Hopefully some of the
suggestions made above will inspire new experiments. Maybe this is the
right moment to give a personal perspective on why one could still be
interested in complex dynamics in one-dimensional systems.

There is no overall applicable principle or method to describe
non-equilibrium systems, but a large subclass is characterized by a
combination of nonlinearity and spatial extend. While in general it is
not clear how to go from the tools developed for low dimensional
dynamical systems to an effective description for systems with many
degrees of freedom, the coherent/incoherent structures framework
sketched in this paper seems to be able to form such bridge in the
case of the CGLE.

Possibly the greatest advantage of studying one-dimensional systems is
that their time evolution can be captured in two-dimensional
space-time plots, which allow the development of intuition for these
systems. Without these, the discovery of most of the Homoclon and MAW
dynamical properties would have been much more difficult.

The uncovered mechanisms by which an unstable wave can give rise to
MAWs, or, beyond the saddle-node, to defects, and the existence of
Homoclons that act as separatrices between defect free and defect
developing states, are not trivial. I have some hope that these
mechanisms may prove to be more general, and therefore it will be
extremely interesting to see what happens in experiments.

Finally, when $\varepsilon$ is increased sufficiently, new mechanisms
will start to play a role in the experiments, and the interplay
between these and the ``low $\varepsilon$'' dynamics described here
will be interesting. The range of experiments, although potentially
large, over which the CGLE dynamics is applicable has not yet been
sufficiently mapped out: let us hope that more will be known in the
next few years.

\subsection{Acknowledgments}
The work to uncover the phase gradient structures and their behavior
was carried out in collaboration with Markus Baer, Lutz Brusch, Martin
Howard, Mads Ipsen, Alessandro Torcini and Martin Zimmerman and I
thank them for an interesting journey. In addition, numerous
discussions with Igor Aranson, Tomas Bohr, Hugues Chat\'e, Francois
Daviaud, Lorenz Kramer, Wim van Saarloos and Willem van de Water are
gratefully acknowledged.

\begin{appendix}

\section{Coefficients of the  CGLE in the labframe}
\label{app}

The spacetime diagrams one obtains in experimental situations cannot
be directly compared with the predictions from the CGLE, since the
simple form of the CGLE (\ref{cgle}) is only obtained after a number
of rescalings and coordinate transforms that I will discuss here.  It
is convenient to introduce first the dimensionless parameter
$\varepsilon$ that measures the distance of the controlparameter to
threshold \cite{ch}. At threshold ($\varepsilon \!=\! 0$), the mode
with dimensional wavenumber $q_c$ and frequency $\omega_c$ (nonzero)
becomes unstable.

The central observation underlying the amplitude equation approach is
that for small but finite positive values of $\varepsilon$ one expects
a {\em band} of wavenumbers of width $\propto \sqrt{\varepsilon}$ to
play a role.  Hence close to threshold one expects that a physical
field $u$ can be written as the product of a slowly varying amplitude
$A$ with the critical mode: $u(x_d,t_d)\propto \exp(i(q_c x_d -
\omega_c t_d)) A( x,t) + c.c.$ \cite{ch}, where $x$ and $t$ denote the
slow non-dimensionalized space and time coordinates. After
substituting this ansatz into the underlying physical equations of
motion (assuming that they are known) and performing a rather tedious
expansion up to third order in $\varepsilon$ \cite{ch,kuo,thesis}, one
then obtains the appropriate amplitude equations: the cubic complex
Ginzburg Landau equation. In its full dimensional form, which is most
appropriate for highlighting the problems one may encounter when
comparing to real data, the equation reads:
\begin{equation}
\tau_0 (\frac{\partial A_d}{\partial t_d} - s_d \frac{\partial
A_d}{\partial x_d}) = \varepsilon (1+ i c_0) A_d + \xi_0^2 (1+ i c_1)
\frac{\partial^2 A_d}{\partial x_d^2} - g_0 (1- i c_3) |A_d|^2 A_d~.
\label{cgle_full}
\end{equation}

To compare experimental data to the CGLE for $A_d$
(Eq. (\ref{cgle_full})), one has to eliminate the fast scales
corresponding to the critical mode, and to do so one proceeds as
follows.
\begin{itemize}
\item Perform a Laplace transform on the experimental spacetime data
set $u$ to obtain a complex valued field $U$ as a function of the
labframe coordinates $x_d$ and $t_d$ \cite{wvdw}.
\item Find the onset for pattern formation, and measure the critical
wavenumber $q_c$ and frequency $\omega_c$, i.e., characterize the wave
obtained for $\varepsilon$ as close to zero as possible.
\item Demodulate the field $U$ to obtain the dimensional field $A_d$,
by writing $A_d \!=\! U \exp(i(q_c x_d - \omega_c t_d))$ (assuming one
wave with positive phase-velocity).
\end{itemize}
One now has obtained a complex field $A_d$ that is described by the
dimensional amplitude equation Eq. (\ref{cgle_full}).

For a theoretical analysis, most of the coefficients occurring in
Eq. (\ref{cgle_full}) can be scaled out. The coefficients $\tau_0$ and
$\xi_0$ give typical temporal and spatial scales to the equation as
given by the dispersion relation for plane waves.  Going to the
so-called dimensionless slow scales $x$ and $t$, which are defined as
$x:= \sqrt{\varepsilon} \xi_0^{-1} x_d$ and $t:= \varepsilon
\tau_0^{-1} t_d$ sets the coefficients $\xi_0, \tau_0$ and the linear
growthrate $\varepsilon$ equal to 1.  By writing $A_d = g_0^{-1/2}
\exp{( i c_0 t)} A$, the coefficients $g_0$ and $c_0$ are scaled to 1
and 0 respectively, and the non-dimensionalized form of $A$ is
obtained. The linear, dimensional group velocity $s_d$ can be removed
by going to a comoving frame, but this only makes sense when the
system has periodic boundary conditions; for a finite system with
fixed boundaries, the value of $s_d$ does play an important role.
Assuming that one can go to the comoving frame, the CGLE in its simple
form (\ref{cgle}) is obtained.

\subsection{Measuring the CGLE coefficients}\label{app_coef}

The coefficients $\tau_0, s_d, c_0, \xi_0^2, c_1, g_0$ and $c_3$ can
in principle be calculated from the underlying equations of motion
\cite{ch,kuo,thesis}. For many experimental situations, however, such
calculations are not available or can only be done in certain
approximations; for some systems the equations of motion or boundary
conditions are not even known in enough detail to allow such
calculations. Since the scale and even qualitative character of the
dynamics depends on these coefficients, their knowledge is
essential. The task of measuring these coefficients appears rather
nightmarish at first sight, but as I hope to show below, may in fact
not be terribly complicated. For recent examples of experimental
determinations of these coefficients see \cite{burg,wvdw,liu}.

\subsubsection{Homogeneous solutions and dispersion relation}
When one searches for homogeneous but possibly growing or decaying
plane wave solutions of Eq. (\ref{cgle_full}) of the form $a(t_d)
\exp(i(q_d x_x - \omega_d t_d))$\footnote{Note that the critical
wavenumber and frequency have already been split off: the bare
labframe wavenumber $q_{lab}$ is equal to $q_c + q_d$.} one obtains a
complex valued equation, of which the real part reads:
\begin{equation}
\partial a /\partial t_d = \varepsilon - \xi_0^2 q_d^2 - g_0
a^2~. \label{d1}\\
\end{equation}
When $a$ is time-independent, the real and imaginary part of this
equation are:
\begin{eqnarray}
\tau_0 \omega &=& \varepsilon c_0 - \tau_0 s_d q_d + \xi_0^2 c_1 q^2 -
g_0 c_3 a^2 \label{d2}\\ 0&=& \varepsilon - \xi_0^2 q_d^2 - g_0
a^2~. \label{d3}
\end{eqnarray}
Finally, when one restricts oneself to plane wave solutions, where Eq.
(\ref{d3}) is satisfied, Eq. (\ref{d2}) becomes the ``nonlinear''
dispersion relation for plane waves:
\begin{equation}
\tau_0 \omega = \varepsilon (c_0-c_3) - \tau_0 s_d q_d + \xi_0^2
 (c_1 +c_3) q^2
\label{d4}
\end{equation}
These equations will be the basis for the experiments described below.

\subsubsection{Quenches of $\varepsilon$}
The coefficients $\tau_0$, $c_0$ and $c_3$ can be obtained by
performing experiments where $\varepsilon$ is suddenly changed from
one value to another. The simplest case is when initially the system
is below threshold $(\varepsilon<0)$ and then suddenly the control
parameter is changed to a positive value $\varepsilon_i$.  A nonlinear
plane wave state will start to grow then.  Assuming that one is in the
linear regime, that this growing wave is spatially homogeneous and has
wavenumber $q_d =0$, one finds from Eqs. (\ref{d1}) that $a \propto
\exp(\tau_0 \varepsilon t_d)$. Denote the time interval during which
the wave goes from 10\% to 50\% of its final strength by $t_i$.
Plotting $\varepsilon$ versus $1/t_i$ then should show a linear
relationship with a slope give by $\tau_0$.

During such quenches one can measure the frequency $\omega$, both when
the wave just starts to grow and when it is fully developed.  From
Eq. (\ref{d2}) one finds that, for $q_d =0$, an infinitesimal wave has
frequency $ \varepsilon/\tau_0 c_0$, while a fully developed wave has
frequency (Eq. (\ref{d4}) $\varepsilon/\tau_0 (c_0-c_3)$. So from
these two measurements $c_0$ and $c_3$ can be determined.

Of course, the assumption that the waves are homogeneous and have
wavenumber $q_d=0$ may not always be satisfied. However, it is known
that for small but positive $\varepsilon$ the band of allowed
wavenumbers shrinks $\propto \sqrt{\varepsilon}$, while the spatial
modulational scales similarly. Therefore, when the growth-rate is
rapidly changed from $\varepsilon_1$ to $\varepsilon_2$ and back a
number of times, where $\varepsilon_1$ is close to zero while
$\varepsilon_2$ is not, one expects to indeed have a homogeneous plane
wave of wavenumber very close to zero. Then one can use Eq. (\ref{d1})
which, for $q_d=0$, becomes $\partial a /\partial t_d = \varepsilon -
g_0 a^2$ to get a full numerical prediction of $a$ as a function of
time. Comparing this full solution to the experimentally obtained
curves for $a(t_d)$, both for the ``up'' as well as for the ``down''
quench for a range of values of $\varepsilon_1$ and $\varepsilon_2$
then gives accurate estimates for both $g_0$ and $\tau_0$.

\subsubsection{Propagation of linear perturbations}
Suppose one has generated a stable plane wave and locally perturbs
this wave. Then, as discussed above, the temporal evolution of this
perturbation will be a combination of slow diffusion and advection
with group velocity $\partial \omega_d / \partial q_d$. For waves of
wavenumber $q_d=0$, this group velocity is the {\em linear} group
velocity $s_d$, and for more general waves one finds by
differentiating the nonlinear dispersion relation that the nonlinear
groupvelocity is $s_d + 2 \xi_0^2 q_d (c_1 + c_3)/\tau_0$.  For a
measurement of $s_d$ it is therefore easiest to make $q_d$ equal to
zero, which can, for example, be done by varying $\varepsilon$ up and
down as described above. It should be noted that while small
perturbations may be difficult to observe in noisy snapshots of the
system, they often become quite clear in space-time plots of the raw
data $u$.

When the perturbations are relatively smooth, such that $\partial a
/\partial t$ is small, a comparison of the instantaneous amplitude $a$
and local wavenumber $q$ leads to another condition on the
coefficients of the CGLE. From Eq. (\ref{d2}) one finds $g_0 |A|^2 =
\varepsilon -\xi_0^2 q^2$. Plotting $|A|^2/\varepsilon$ versus
$q^2/\varepsilon$ one expects a linear relation. For $q^2 \downarrow
0$ the value of $|A|^2/\varepsilon$ approaches $g_0$, while the slope
of the curve will be equal to $\xi_0^2/g_0$.

\subsubsection{Fronts}

When $\varepsilon$ is suddenly increased from below to above
threshold, details of the noise and the initial conditions determine
the details of the growth of the nonlinear state. Above it has been
assumed that one has a homogeneous state, but when a localized
perturbation is applied just before such a change of $\varepsilon$
occurs, it is possible to create a pair of fronts that propagate into
the unstable $a=0$ state, and leave a nonlinear state in their
wake. The point is that for large times the velocity of the front and
the wavenumber of the nonlinear state are selected and can be
calculated from an essentially linear analysis \cite{ch,saar1}.  These
velocities are $s_d \pm 2 \xi_0/\tau_0 \sqrt{\varepsilon(1+c_1^2)}$,
and the wavenumber of the plane wave they leave behind is $q_c +
\xi_0^{-1} \sqrt{\varepsilon/(1+ c_1^2)}$ \cite{saar1}.  It should be
noted that the velocity and waveumber only relax to their asymptotic
values as $1/t$, so it is hard to avoid making some errors here.

\subsubsection{Eckhaus instability}

As discussed in the section on MAWs, one can manipulate the effective
wavenumber of a plane wave state by changes of $\varepsilon$. At some
point, when the effective value of $q$ becomes too large, one will
encounter the Eckhaus instability, which for general $\varepsilon$ is
given by $q^2 < \varepsilon \xi_0^{-2}
\frac{(1-c_1c_3)}{3-c_1c_3+2c_3^2}$ \cite{bf}. This could be used as
an independent check on the coefficients.

\section{Coherent structure ODEs}\label{app2}

The coherent structure ODEs obtained when the ansatz
Eq. (\ref{coh_ans}) is substituted into the CGLE can be written in a
number of forms and for a number of different dependent variables. All
useful representations will have $a$ and $q:= \partial_{\xi} \phi$ as
variables, but there are broadly speaking two choices; either
$b:=\partial_{\xi} a $ or $\kappa:=(\partial_{\xi} a) /a$
\cite{saar1,clon,maw,lutznew}.  The latter representation is
particularly useful if one wants to study structures, such as fronts,
that asymptotically decay to $A=0$.  $\kappa$ then measures the
exponential decay of the profile to the zero state.  Using this latter
notation the set of coupled ordinary differential equations (ODE's)
becomes
\begin{eqnarray}\label{ode1}
\partial_{\xi} a &=& \kappa a ~,  \\
 \partial_{\xi} z
&=& -z^2 +\frac{1}{1+i c_1}\left[ - 1 - i \omega + (1-i c_3)a^2
-v z\right] ~,\label{ode2}
\end{eqnarray} \noindent
where the complex quantity $z$ is equal to $\kappa + i q$, which is
also equal to $\partial_{\xi} \ln(A)$.  Equation (\ref{ode2}) is
equivalent to two real valued equations, so (\ref{ode1},\ref{ode2})
can also be seen as a 3D real-valued dynamical system \cite{saar1}.
Note that since the CGLE is a complex, second order equation one may
have expected to find 4 coupled ODEs, but they can be reduced to 3
equations using the phase symmetry of the CGLE.

For completeness note that another ansatz is frequently encountered in
the literature: $A= a(\xi) \exp( i (q x - \omega t)) \exp(i \phi(\xi))
$. This may be convenient if one wants to split off an explicit
wavenumber $q$ and require that $\partial_{\xi} \phi$ goes to zero for
$\xi \rightarrow \pm \infty$. It is easy to show that this ansatz is
equivalent to ansatz (\ref{coh_ans}), and so the apparent three free
parameters of this ansatz can be reduced to two \cite{review}.

\end{appendix}

\bibliographystyle{elsart-num}

\begin{thebibliography}{10}

\bibitem{nb} K. Nozaki and N. Bekki, Jour. Phys. Soc. Jap. {\bf 53},
1581 (1984).

\bibitem{maw} L. Brusch, M. G. Zimmermann, M. van Hecke, M. B{\"a}r,
A. Torcini, Phys. Rev.  Lett. {\bf85}, 86 (2000); L. Brusch,
A. Torcini, M. van Hecke, M. G. Zimmermann and M. B{\"a}r, Physica D
{\bf 160}, 127 (2001).
 
\bibitem{lutznew} L. Brusch, A. Torcini and M. B\"ar, preprint (2001);
L. Brusch, ``Complex patterns in extended oscillatory systems'', PhD
thesis, Dresden (2001).

\bibitem{clon} M. van Hecke, Phys. Rev. Lett.  {\bf 80}, 1896 (1998).

\bibitem{mm} M. van Hecke, M. Howard, Phys. Rev. Lett. {\bf 86}, 2018
(2001); M. Howard and M. van Hecke, in preparation.

\bibitem{zz} M. Ipsen and M. van Hecke, Physica D {\bf 160}, 103
(2001).

\bibitem{nbunst} S. Popp, O. Stiller, I. Aranson, A. Weber and
L. Kramer, Phys Rev Lett {\bf70}, 3880 (1993); S. Popp, O. Stiller,
I. Aranson and L. Kramer, Physica D {\bf 84}, 398 (1995); O. Stiller,
S. Popp and L. Kramer, Physica D {\bf84}, 424 (1995) ; O. Stiller,
S. Popp, I. Aranson and L. Kramer, Physica D {\bf87}, 361 (1995);

\bibitem{chate} H. Chat{\'e}, Physica D {\bf86}, 238 (1995).

\bibitem{spiral} I. S. Aronson, L. Aronson, L. Kramer and A. Weber,
Phys. Rev. A {\bf46}, R2992 (1992); G. Huber, P. Ahlstroem and
T. Bohr, Phys. Rev. Lett. {\bf 69}, 2380 (1992).

\bibitem{nbint} H. Sakaguchi, Prog. Theor. Phys. {\bf85}, 417 (1991);
H. Sakaguchi, Prog. Theor. Phys. {\bf 86}, 7 (1991); H. Chat\'e and
P. Manneville, Phys. Lett. A {\bf 171}, 183 (1992).

\bibitem{review} L. Kramer and I. Aranson, Rev. Mod. Phys. {\bf 74},
99 (2002).

\bibitem{lega} J. Lega, B. Janiaud, S, Jucquois and V. Croquette,
Phys. Rev. A {\bf 45}, 5596 (1992).

\bibitem{fleselles} J.-M. Flesselles, V. Croquette and S. Jucquois,
Phys. Rev. Lett. {\bf 72}, 2871 (1994).

\bibitem{burg} J. Burguete, H. Chat\'e, F. Daviaud and N. Mukolobwiez,
Phys. Rev. Lett. {\bf 82}, 352 (1999).

\bibitem{ch} M. C. Cross, P. C. Hohenberg, Pattern formation outside
of equilibrium, Rev. Mod. Phys. {\bf65}, 851 (1993).

\bibitem{mysource}M. van Hecke, C. Storm, W. van Saarloos, Physica D
134 (1999) 1 and references therein

\bibitem{source} P. Coullet, T. Frisch and F. Plaza, Physica D {\bf
62}, 75 (1993); E. Knobloch and J. Deluca, Nonlinearity {\bf 3}, 975
(1990); C. Martel and J. M. Vega, Nonlinearity {\bf 11}, 105 (1998); H. Riecke and L. Kramer, Physica D {\bf 137}, 124
(2000).

\bibitem{shrai} B. I. Shraiman, A. Pumir, W. van Saarloos,
P. C. Hohenberg, H. Chat{\'e}, M. Holen, Physica D {\bf 57}, 241
(1992).

\bibitem{chat2} H. Chat{\'e}, Nonlinearity {\bf 7} 185  (1994). 

\bibitem{bf} T. B. Bejamin and J. E. Feir, J. Fluid Mech.  {\bf 27},
417 (1967).

\bibitem{janiaud} B. Janiaud, A. Pumir, D. Bensimon, V. Croquette,
H. Richter and L. Kramer, Physica D {\bf55}, 269 (1992).

\bibitem{kramer85} L. Kramer and W. Zimmermann, Physica D {\bf16}, 221
(1985).

\bibitem{wound} R. Montagne, E. Hernandez-Garcia and M. San-Miguel,
Phys. Rev. Lett. {\bf77}, 267 (1996); A. Torcini,
Phys. Rev. Lett. {\bf77} 1047 (1996).


\bibitem{saar1} W. van Saarloos, P. C. Hohenberg, Physica D {\bf56},
303 (1992); Physica D {\bf69}, 209 (1993) (errata).

\bibitem{chif} N. Mukolobwiez, A. Chiffaudel and F. Daviaud,
Phys. Rev. Lett. {\bf 80}, 4661 (1998). 

\bibitem{liu} Y. M. Liu and R. E Ecke, Phys. Rev.  E {\bf59}, 4091
(1999).

\bibitem{53lutz} Y. Liu and R. E. Ecke, Phys. Rev. Lett. {\bf 78},
4391 (1997).


\bibitem{hw} J. M. Vince and M. Dubois, Physica D {\bf 102}, 93
(1997).

\bibitem{wvdw} L. Pastur, M.-T. Westra, W. van de Water, M. van Hecke,
C. Storm and W. van Saarloos, arXiv cond-mat/0111234; L. Pastur,
M.-T. Westra and W. van de Water, Physica D, this issue.

\bibitem{garnier} N. Garnier and A. Chiffaudel, Phys Rev. Lett. {\bf
86}, 75 (2001); idem, this issue of Phys. D. 

\bibitem{kroket} V. Croquette and H. Williams, Phys. Rev. A {\bf39},
2765 (1989). 

\bibitem{bot2} P. Bot, O. Cadot and I. Mutabazi, Phys. Rev. E {\bf
58}, 3089 (1998).
 


\bibitem{privcom} J.-M. Flesselles, private communications (1999).

\bibitem{bot} P. Bot and I Mutabazi, Eur. Phys. J. B. {\bf 13}, 141
(2000).

\bibitem{kuo} E. Y. Kuo and M. C. Cross, Phys. Rev. E {\bf47}, R2245
(1993).

\bibitem{thesis} M. van Hecke and W. van Saarloos,
Phys. Rev. E. {\bf55}, R1259 (1997); M. van Hecke, 'The amplitude
description of nonlinear patterns', PhD thesis, Leiden University
(1996).

\end{thebibliography}

\end{document}